
\newbox\leftpage \newdimen\fullhsize \newdimen\hstitle \newdimen\hsbody
\tolerance=1000\hfuzz=2pt
\magnification=1200\baselineskip=16pt plus 2pt minus 1pt
\hsbody=\hsize \hstitle=\hsize 
\catcode`\@=11 
\newcount\yearltd\yearltd=\year\advance\yearltd by -1900

%
%
\def\draftmode{\def\draftdate{{\rm preliminary draft:
\number\month/\number\day/\number\yearltd\ \ \hourmin}}%
\headline={\hfil\draftdate}\writelabels\baselineskip=20pt plus 2pt minus 2pt
{\count255=\time\divide\count255 by 60 \xdef\hourmin{\number\count255}
        \multiply\count255 by-60\advance\count255 by\time
   \xdef\hourmin{\hourmin:\ifnum\count255<10 0\fi\the\count255}}}

\def\nolabels{\def\eqnlabel##1{}\def\eqlabel##1{}\def\reflabel##1{}}
\def\writelabels{\def\eqnlabel##1{%
{\escapechar=` \hfill\rlap{\hskip.09in\string##1}}}%
\def\eqlabel##1{{\escapechar=` \rlap{\hskip.09in\string##1}}}%
\def\reflabel##1{\noexpand\llap{\string\string\string##1\hskip.31in}}}
\nolabels
%
\global\newcount\secno \global\secno=0
\global\newcount\meqno \global\meqno=1
\def\newsec#1{\global\advance\secno by1\message{(\the\secno. #1)}
\xdef\secsym{\the\secno.}\global\meqno=1
\bigbreak\bigskip
\noindent{\bf\the\secno. #1}\par\nobreak\medskip\nobreak}
\xdef\secsym{}
\def\appendix#1#2{\global\meqno=1\xdef\secsym{\hbox{#1.}}\bigbreak\bigskip
\noindent{\bf Appendix #1. #2}\par\nobreak\medskip\nobreak}
%
%
\def\eqnn#1{\xdef #1{(\secsym\the\meqno)}%
\global\advance\meqno by1\eqnlabel#1}
\def\eqna#1{\xdef #1##1{\hbox{$(\secsym\the\meqno##1)$}}%
\global\advance\meqno by1\eqnlabel{#1$\{\}$}}
\def\eqn#1#2{\xdef #1{(\secsym\the\meqno)}\global\advance\meqno by1%
$$#2\eqno#1\eqlabel#1$$}
%
\newskip\footskip\footskip14pt plus 1pt minus 1pt 
\def\f@@t{\baselineskip\footskip\bgroup\aftergroup\@foot\let\next}
\setbox\strutbox=\hbox{\vrule height9.5pt depth4.5pt width0pt}
\global\newcount\ftno \global\ftno=0
\def\foot{\global\advance\ftno by1\footnote{$^{\the\ftno}$}}
%
%
\global\newcount\refno \global\refno=1
\newwrite\rfile
\def\ref{\nref}
\def\nref#1{\xdef#1{[\the\refno]}\ifnum\refno=1\immediate
\openout\rfile=refs.tmp\fi\global\advance\refno by1\chardef\wfile=\rfile
\immediate\write\rfile{\noexpand\item{#1\ }\reflabel{#1}\pctsign}\findarg}
\def\findarg#1#{\begingroup\obeylines\newlinechar=`\^^M\pass@rg}
{\obeylines\gdef\pass@rg#1{\writ@line\relax #1^^M\hbox{}^^M}%
\gdef\writ@line#1^^M{\expandafter\toks0\expandafter{\striprel@x #1}%
\edef\next{\the\toks0}\ifx\next\em@rk\let\next=\endgroup\else\ifx\next\empty%
\else\immediate\write\wfile{\the\toks0}\fi\let\next=\writ@line\fi\next\relax}}
\def\striprel@x#1{} \def\em@rk{\hbox{}} {\catcode`\%=12\xdef\pctsign{
\def\semi{;\hfil\break}
\def\addref#1{\immediate\write\rfile{\noexpand\item{}#1}} 
\def\listrefs{\immediate\closeout\rfile
\baselineskip=14pt\centerline{{\bf References}}\bigskip{\frenchspacing%
\escapechar=` \input refs.tmp\vfill\eject}\nonfrenchspacing}
\def\startrefs#1{\immediate\openout\rfile=refs.tmp\refno=#1}
\def\figures{\centerline{{\bf Figure Captions}}\medskip\parindent=40pt}
\def\fig#1#2{\medskip\item{Fig.~#1:  }#2}
\catcode`\@=12 
%
%
\def\noblackbox{\overfullrule=0pt}
\hyphenation{anom-aly anom-alies coun-ter-term coun-ter-terms}
\def\inv{^{\raise.15ex\hbox{${\scriptscriptstyle -}$}\kern-.05em 1}}
\def\dup{^{\vphantom{1}}}
\def\Dsl{\,\raise.15ex\hbox{/}\mkern-13.5mu D} 
\def\dsl{\raise.15ex\hbox{/}\kern-.57em\partial}
\def\del{\partial}
\def\Psl{\dsl}
\def\tr{{\rm tr}} \def\Tr{{\rm Tr}}
\def\lspace{\ifx\answ\bigans{}\else\qquad\fi}
\def\lbspace{\ifx\answ\bigans{}\else\hskip-.2in\fi} 
\def\boxeqn#1{\vcenter{\vbox{\hrule\hbox{\vrule\kern3pt\vbox{\kern3pt
        \hbox{${\displaystyle #1}$}\kern3pt}\kern3pt\vrule}\hrule}}}
\def\mbox#1#2{\vcenter{\hrule \hbox{\vrule height#2in
                \kern#1in \vrule} \hrule}}  
%
\def\CAG{{\cal A/\cal G}}   
\def\CA{{\cal A}} \def\CC{{\cal C}} \def\CF{{\cal F}} \def\CG{{\cal G}}
\def\CL{{\cal L}} \def\CH{{\cal H}} \def\CI{{\cal I}} \def\CU{{\cal U}}
\def\CB{{\cal B}} \def\CR{{\cal R}} \def\CD{{\cal D}} \def\CT{{\cal T}}
\def\e#1{{\rm e}^{^{\textstyle#1}}}
\def\grad#1{\,\nabla\!_{{#1}}\,}
\def\gradgrad#1#2{\,\nabla\!_{{#1}}\nabla\!_{{#2}}\,}
\def\ph{\varphi}
\def\psibar{\overline\psi}
\def\om#1#2{\omega^{#1}{}_{#2}}
\def\vev#1{\langle #1 \rangle}
\def\lform{\hbox{$\sqcup$}\llap{\hbox{$\sqcap$}}}
\def\darr#1{\raise1.5ex\hbox{$\leftrightarrow$}\mkern-16.5mu #1}
\def\lie{\hbox{\it\$}} 
\def\ha{{1\over2}}
\def\half{{\textstyle{1\over2}}} 
\def\roughly#1{\raise.3ex\hbox{$#1$\kern-.75em\lower1ex\hbox{$\sim$}}}
%
\parskip=5 pt plus 3pt minus 1pt
\global\newcount\parno \global\parno=0
\def\newpar#1{\global\advance\parno by1 \bigbreak \bigskip
\the\secno-\the\parno. #1 \par \nobreak \bigskip \nobreak}
\pageno=0
\footline={\ifnum\pageno=0\else\hfil\number\pageno\hfil\fi}
\def\of#1{\left( #1 \right)}
\def\abs#1{ \left\vert #1 \right\vert}
\def\at#1#2{ \left. #1 \right\vert_{#2}}
\def\eps{\varepsilon}
\def\repart{ \Re {\rm e}}
\def\impart{Im }
\def\i{\sf i}

\def\ph{\varphi}

\ref\zamcth{A.B.Zamolodchikov, JETP Lett. {\bf 43} (1986) 730;
Sov.J.Nucl.Phys. {\bf 46} (1987) 1090.}

\ref\shenker{D.A.Kastor, E.J.Martinec and S.H.Shenker, Nucl. Phys.
{\bf B316} (1989) 590.}

\ref\cardycth{J.L.Cardy, Phys.Lett. {\bf B215} (1988) 749.}

\ref\osborn{H.Osborn, Phys.Lett. {\bf B222} (1989) 97 \semi
    I.Jack and H.Osborn, Nucl. Phys. {\bf B343} (1990) 647.}

\ref\mavro{N.E.Mavromatos, J.L.Miramontes and J.L.S\'anchez de Santos,
Phys.Rev. {\bf D40} (1989) 535.}

\ref\shore{G. Shore, Phys.Lett. {\bf B253} (1991) 380 ; {\bf B256} (1991) 407.}

\ref\cfl{A. Cappelli, D. Friedan and J.I.Latorre, Nucl.Phys.
{\bf B352} (1991) 616.}

\ref\damgaard{P.H.Damgaard, {\it Stability and Instability of the
Renormalization Group Flows}, preprint CERN-TH-6073-91.}

\ref\ludwig{A.W.Ludwig and J.L.Cardy, Nucl.Phys. {\bf B285} (1987) 687.}

\ref\lawrie{C.Athorne and I.D.Lawrie, Nucl.Phys. {\bf B265} (1985) 577.}

\ref\wilson{K.G.Wilson and J.Kogut, Phys.Rep. {\bf 12C} (1974) 75.}

\ref\wz{D.J.Wallace and R.K.P.Zia, Ann. Phys. (NY) {\bf 92} (1975) 142.}

\ref\lipatov{L.N.Lipatov, Sov. Phys. JETP {\bf 44} (1976) 1055.}

\ref\brezin{E.Br\'ezin, J.C.Le Guillou and J.Zinn-Justin, in
{\it Phase Transitions and Critical Phenomena, vol.6}, ed.
C.Domb and M.S.Green (Academic Press, 1976).}

\ref\zamlg{A.B.Zamolodchikov, Sov.J.Nucl.Phys. {\bf 44} (1986) 529.}

\ref\zinn{J.Zinn-Justin, {\it Quantum field theory and critical
phenomena} (Clarendon, Oxford 1989)\semi
E.Br\'ezin and J.Zinn-Justin, Phys. Rev. {\bf B14} (1976) 3110.}

\ref\cl{A.Cappelli and J.I.Latorre, Nucl.Phys. {\bf B340} (1990) 659.}

\ref\flv {D.Z. Freedman, J.I.Latorre and X. Vilas\'{\char'20}s,
Mod.Phys.Lett. {\bf A6} (1991) 531.}

\ref\cardysumrule{J.L.Cardy, Phys.Rev.Lett. {\bf 60} (1988) 2709.}

\ref\cardyshear{J.L.Cardy, Nucl.Phys. {\bf B290} (1987) 355 and preprint
UCSB-90-38\semi
A.Cappelli and A.Coste, Nucl.Phys. {\bf B314} (1989) 707.}

\ref\bpz{A.A.Belavin, A.M.Polyakov and A.B.Zamolodchikov, Nucl.Phys.
{\bf B241} (1984) 333.}

\ref\vafa{C.Vafa, Phys.Lett. {\bf B212} (1988) 28.}

\ref\lassig{M.L\"assig, Nucl. Phys. {\bf B334} (1990) 652.}

\ref\itzykson{C.Itzykson and J.-M.Drouffe, {\it Statistical Field Theory}
(Cambridge Univ. Press, 1989).}

\ref\felder{G.Felder, Com. Math. Phys.{\bf 111} (1987) 101.}

\ref\on{B.Rosenstein, B.J.Warr and S.H.Park, Nucl.Phys. {\bf B336}
(1990) 435.}

\ref\curci{G.Curci, G.Paffuti, Nucl.Phys. {\bf B312} (1989) 227.}


\line{\hfill CERN-TH-6201/91}
\line{\hfill August 1991}

\vskip .6truein
\centerline{\bf RENORMALIZATION GROUP PATTERNS AND C-THEOREM}
\centerline{\bf IN MORE THAN TWO DIMENSIONS}
\vskip .5 truein
\centerline{Andrea CAPPELLI\footnote*{On leave of absence from INFN,
Sezione di Firenze, Italy. Bitnet: cappelli@cernvm.},
Jos\'e Ignacio  LATORRE\footnote{**}{On leave of absence
from Departament ECM, Barcelona. Bitnet: latorre@ebubecm1.} }
\centerline
{\it Theory Division, CERN, Geneva, Switzerland}
\bigskip
\bigskip
\centerline{and}
\bigskip
\centerline{ Xavier VILAS\'IS-CARDONA\footnote{***} {Bitnet: druida@ebubecm1.}}
\centerline{\it Departament d'Estructura i Constituents de la Mat\`eria}
\centerline{\it University of Barcelona}
\centerline{\it Av. Diagonal 647, 08028 Barcelona, Spain}
\bigskip\bigskip\bigskip

\centerline{ABSTRACT}
We elaborate on a previous attempt to prove the irreversibility of the
renormalization group flow above two dimensions.
This involves the construction of a monotonically decreasing $c$-function
using a spectral representation.
The missing step of the proof is a good definition of this function
at the fixed points.
We argue that for all kinds of perturbative flows the $c$-function is
well-defined and the $c$-theorem holds in any dimension.
We provide examples in multicritical and multicomponent
scalar theories for dimension $2<d<4$.
We also discuss the non-perturbative flows in the yet unsettled
case of the $O(N)$ sigma-model for $2\leq d\leq 4$ and large $N$.

\vbox to .6 truein {\vfill}
\vfill
\line{CERN-TH-6201/91\hfill }
\line{August 1991\hfill }
\eject

\newsec{Introduction}
\bigskip\bigskip
There  is a common belief which says that the
renormalization group (RG) flows are irreversible.
Intuitively,  short-distance degrees of freedom are
integrated out in order to obtain  a long-distance
effective description of a physical system and
are, therefore,  irrecoverable. Materializing this intuition
into a theorem has proven to be quite a hard  task, so far unfinished for
dimensions $d>2$.

Zamolodchikov produced a theorem in two dimensions based on
the explicit construction of a function which decreases monotonically
along RG trajectories \zamcth. This is obtained from the
two-point function of the stress tensor and it is called
``$c$-function'' because it coincides  at fixed
points with the central charge of the corresponding conformal field theory.
In his original proof, Zamolodchikov assumed Poincar\'e invariance, locality,
renormalizability and, notably, unitarity. Actually, the
very irreversibility of the RG flow, {\sl i.e.} the monotonicity of the
$c$-function, is due to unitarity.

The $c$-theorem is by now a valuable tool for non-perturbative field
theory in two dimensions. A striking example is the proof
of spontaneous breaking of supersymmetry in the flow from the
tricritical to the critical Ising model \shenker.
A similar tool would be very interesting in higher dimensions,
to investigate long-standing non-perturbative problems like confinement,
chiral symmetry breaking and the Higgs mechanism.
The explicit ingredients in Zamolodchikov's proof are not specific
to two dimensions, but so far problems have been found to extend
it to higher dimensions.

The efforts to enlarge the validity of the theorem can be roughly
divided in two classes.
In the first one, some $c$-numbers characteristic of four-dimensional
critical theories are studied, mainly those parametrizing the
gravitational trace anomaly \cardycth\osborn\footnote*{
See also the approach of ref.\mavro.}.
The ingredient of unitarity is not used, so that the
monotonicity of these quantities along the RG flow cannot be proven.
An effort to use positivity along these lines
was done in ref.\shore.

In the second approach, the constraint of unitarity is explicitly built in.
In ref. \cfl, a refined version of the two-dimensional theorem,
originally due to Friedan, was obtained using the Lehmann spectral
representation for the correlator of two stress tensors.
The $c$-function is issued from data of the Hilbert space of the
theory, so that renormalization problems are bypassed.
This approach can be extended to higher dimensions, except for
one point, the meaning of the $c$-function at the fixed points is
unclear. Thus, we cannot  claim that the theorem is conquered yet.

Known counterexamples of RG flows with complex behaviour, like
limit cycles and chaos, have been discussed in ref. \damgaard.
They violate some of the assumptions of the theorem, namely Poincar\'e
invariance (spin glasses, hierarchical models) or unitarity
(polymers, models with replica trick\footnote*
{They are not unitarity in two dimensions \ludwig~ and likely
not in any dimension \lawrie.}).

In this paper, we intend  to present the state of the art of the
spectral approach to the $c$-theorem in more than two dimensions.
In sect.  2, we recall its main points.
We stress that the $c$-function is actually well-defined for
perturbative flows, where it changes infinitesimally between
the ultraviolet (UV) and the infrared (IR) fixed points,
$\Delta c = c_{UV} - c_{IR} \ll 1$.

Therefore the RG flow is irreversible in any perturbative expansion.
This fact is of great practical importance, due to the major role of
perturbative expansions in higher dimensions, even if it is far from
the non-perturbative goals we mentioned before.

In sect.  3, we substantiate this claim by studying the $c$-function
in arbitrary dimension using Wilson's epsilon
expansion in scalar theories as a bench-mark \wilson.
In this framework, we compute the RG patterns of the $\lambda \ph^4$
theory \wz, of the multicritical $\lambda \ph^{2r}$ \lipatov~ and of the
multicomponent $g_{ijkl} \ph_i\ph_j\ph_k\ph_l$ theories \brezin.
A typical variation of the $c$-function is
$\Delta c\sim \epsilon^3\ll 1$ in $d=4-\epsilon$.
As expected, the $c$-theorem works perfectly.
It supports the belief that the multicritical pattern of the minimal conformal
models in two dimensions \zamlg~ extends smoothly to higher dimensions.
Moreover, the $c$-functions correctly add up for chains of flows
in the multicomponent theory.

In sect.  4,  we try to understand more difficult flows, which are
``non-perturbative'' in the following sense.
We investigate the $O(N)$ sigma-model in the large $N$ limit for
$2\leq d\leq 4$ \brezin\zinn.
In this case, the expansion is perturbative in $1/N$ (the
coupling of the theory given by the connected 4-point function),
but it is actually non-perturbative for what concerns
the $c$-function. Indeed, the flow in the massive phase gives
$\Delta c\sim N \gg 1$ and is definitely different
from that of free massive bosonic particles.
Our results agree with the known RG pattern of the model for $d=2$ and
$d=4$, but disagree with it for $2<d<4$.
We sketch some possible explanations of this fact which deserve
further investigation.
In the conclusions, we comment on four-dimensional physical theories like QCD.
The Appendix is devoted to setting the technique of
conformal perturbation expansion, valid in any dimension.

\vfill\eject

\newsec{Steps Towards the Proof of the $c$-theorem }

Let us first recall the main features of the $c$-theorem in two
dimensions \cfl.
We define the stress tensor $T_{\mu\nu} (x)$ as the response of the
action to small fluctuations of the spacetime metric.
Then we consider the spectral representation of the Euclidean correlator
\eqn\maintt{
< T_{\mu\nu} (x) T_{\rho\sigma} (0) > ={\pi\over 3}
   \int_0^\infty  d \mu \, c(\mu)
   \int {d^2 p\over (2\pi)^2}\,
      {\rm  e}^{i p x}{
        (  g_{\mu\nu} p^2 -p_\mu p_\nu)
        ( g_{\rho\sigma} p^2 - p_\rho p_\sigma)
        \over p^2 + \mu^2}\quad.}
In this expression, the spectral density $c(\mu)$ is a scalar function
by Poincar\'e invariance and it is positive definite by unitarity,
$c(\mu)\ge 0$.
In the short-distance limit, the theory is described by a UV
conformal invariant theory, where eq. \maintt~ reduces to the single component
$\langle T_{zz}(z) T_{zz}(0)\rangle_{CFT} = c_{UV}/2z^4$,
parametrized by the Virasoro central charge $c_{UV}$.
The same happens in the long-distance (IR) limit.
Working out these two limits in eq. \maintt, it follows that
\eqn\cth{
c_{UV}=\int_0^\infty d\mu\, c(\mu) \quad \quad
\geq \qquad\qquad c_{IR} = \lim_{\epsilon\to 0}
\int_0^\epsilon d\mu\, c(\mu)\qquad.}
where the inequality stems from unitarity.
Thus $c(\mu)d\mu$ is a dimensionless measure of degrees of freedom
off-criticality.

The proof is completed by considering the RG flow of the spectral density.
The previous equation implies that
\eqn\cflow{
c(\mu) = c_{IR} \delta(\mu) + c_1(\mu,\Lambda) \qquad , \qquad
\Delta c\equiv c_{UV} - c_{IR} =\int_0^\infty d\mu \quad c_1 (\mu)\quad.}
The evolution of $c(\mu)d\mu$ under the RG flow is governed by the flow of
the physical mass scale $\Lambda$ of the theory.
As we quit the UV fixed point $\Lambda=0$,
the delta term in eq. \cflow~ stays constant, because it measures the
states of the Hilbert space which remains massless for $\Lambda \neq 0$.
Instead, the states acquiring mass contribute to the smooth density
$c_1(\mu,\Lambda)$, roughly bell-shaped and peaked at $\mu\sim\Lambda$.
In the IR limit $\Lambda\to\infty$, $c_1$ is pushed away and
contributes no more to observables. Thus its integral $\Delta c$ gives a
quantitative measure of the loss of degrees of freedom along the RG flow.

In a typical application of the theorem \cl\flv, $c_{UV}$ and $c_{IR}$ are
first determined from $\langle TT\rangle_{CFT}$ in the corresponding conformal
field theories. Next, the correlator $\langle\Theta\Theta\rangle$
of the interpolating off-critical theory is considered.
By eq. \cflow, the variation $\Delta c$ is independently measured by
the sum rule \cardysumrule
\eqn\realsumrule{
\Delta c ={3\over 4\pi} \int_{\vert x\vert >\epsilon}d^2 x \,x^2
<\Theta(x)\Theta(0)>\quad.}
Thus the data of the critical and off-critical theories are compared
and the RG pattern is verified.

Let us stress some virtues of this proof.
The $c$-theorem expresses a geometrical property of the space of theories,
parametrized by some coupling coordinates $\{g^i\}$.
But notice that our proof was given in a coordinate-free
language. We did not need to talk of bare Lagrangians and renormalization
conditions on fields and couplings, nor  care about their associated
reparametrization invariance.
Actually, the Lehmann representation gives us the spectral density
expressed in terms of matrix elements of the stress tensor, belonging
to the Hilbert space of the theory.
Moreover, these matrix elements are necessarily non-vanishing,
because any matter couples to the stress tensor.
Thus, this is a good measure of degrees of freedom
in the theory. Any other current would not couple to all of them
(depending on their charges) and would not detect their flow.
Finally, the density $c(\mu)$ summarizes all the unitarity conditions
on the two-point function, because any positive quantity
can be obtained by integrating it against positive smearing
functions.

We believe that these are rather unique features, which should
necessarily appear in any generalization of the theorem to higher
dimensions.

For later reference, let us also express the theorem in terms of the
coordinates $\{g^i\}$ and beta-functions
$\Lambda {d\over d\Lambda} g^i=\beta^i (g)$ \zamcth.
One has to expand the trace of the stress tensor $\Theta(x)$ in the
basis of renormalized fields $\Phi_i$ at the UV fixed point,
\eqn\trace{
\Theta(x) = 2\pi \beta^i \Phi_i\quad.}
Next, the $c$-function $c=c(g)$ and the Zamolodchikov metric
$G_{ij}(g)\propto \langle \Phi_i(x)\Phi_j(0)\rangle\vert_{\vert x\vert =1}$
are introduced by smearing $c(\mu)$ against appropriate positive
functions, which contain a fixed scale \cfl.
It follows
\eqn\zcth{
{d\over dt}c\equiv -\beta^i{\partial\over\partial g^i}c(g)=
\,\,- \,\, \beta^i\beta^j G_{ij}(g) \,\, \le 0\quad,}
Thus, the previous flow of $c(\mu)$ driven by a physical mass $\Lambda$
is now traded for the change of $c(g)$ along the flow curve of affine
parameter $t$.

The spectral form of the $c$-theorem can be generalized to higher
dimensions, where it shows a new feature.
There are two spectral densities, $c^{(0)}(\mu)$, related to
spin-zero intermediate states, and  $c^{(2)}(\mu)$ for spin-two  ones.
Both densities are, in principle, candidates for a $c$-theorem,
since they display some of the properties of the unique
density in two dimensions.
$c^{(2)}(\mu)$ determines the correlator of two stress tensors
at the conformal invariant points \cardyshear, and it could define
an analog of the  central charge.
However, by inspection this does not correspond
to a monotonically decreasing function along RG trajectories
and it must be discarded \cfl.

On the other hand, $c^{(0)}(\mu)$ is related
to changes of scale off-criticality, because
\eqn\ththd{
<\Theta(x) \Theta(0)>
=A \,\int^\infty_0 d\mu \,c^{(0)}(\mu) \int
{d^d p\over (2\pi)^d} e^{ipx} {p^4\over p^2+\mu^2} }
\eqn\ththconst{
A={V\over \Gamma(d)(d+1)2^{d-1}}\qquad,\quad
V\equiv Vol(S^{d-1})={2\pi^{d\over2} \over \Gamma({d\over 2})}\quad.}
Therefore, we can generalize the sum rule \realsumrule.
Limiting ourselves to theories with vanishing $\Theta$
at fixed points, {i.e.} conformally invariant fixed points,
a dimensional analysis led us to define
\eqn\dsumrule{
\Delta c= \int_\epsilon^\infty d\mu {c^{(0)}(\mu,\Lambda) \over \mu^{d-2}}
= {d+1\over  V d}\,\int_{\vert x\vert >\epsilon}
 d^dx\, x^d \langle \Theta(x) \Theta(0)\rangle\quad. }
The normalization of $c^{(0)}(\mu)$ in eq. \ththd~ assigns $c=0$ to the
trivial theory and $c=1$ to the free bosonic theory in any dimension,
the latter being computed by the sum rule \dsumrule~ in the free
massive phase.
By smearing $c^{(0)}(\mu)$ we can also generalize eq. \zcth~,
and get a new $c$-function $c(g)$ in $d$ dimensions, which is monotonically
decreasing along the RG flow and stationary at fixed points,
in close analogy with the two-dimensional case \cfl.

However, a point is missing, the characterization of $c(g)$ at fixed points,
the would-be central charge, or $c$-charge.
A closer inspection shows that this is defined as a
limit from off-criticality of the spin-zero density,
\eqn\clim{
\lim_{\Lambda\to 0} {c^{(0)}(\mu,\Lambda)\over \mu^{d-2}}d\mu
= c \,\delta(\mu) d\mu\quad.}
In general, this limit may depend on the path
approaching the fixed point, implying unacceptable non-universal
effects on the $c$-charge.
If $c$ is monotonic but multivalued, we can still have closed
cycles violating the theorem.
On the other hand, the limit is universal if the $c$-charge is
related to an observable of the fixed-point theory
- we could not prove this fact so far.
Note that $\langle\Theta\Theta\rangle_{CFT}=0 $ for $d>2$,
and is thus independent of $c$, owing to the factor $\mu^{d-2}$ in eq. \clim.

As we pointed out above, in two dimensions $c$ has indeed an independent
characterization at the fixed point from $\langle TT\rangle$.
The unique density is both responsible for controlling changes of scale
(genuine spin 0 in $d\geq 2$) and for the coefficient of the short
distance singularity of $T_{zz}(x)T_{zz}(0)$, (genuine spin 2 in $d\geq 2$).
At the conformal point, the trace of the stress tensor still sees
the central charge through contact terms. This is actually
enforced by Lorentz invariance since $T_{zz}$ and $\Theta$ are
just different components of the same Lorentz structure \cfl.
Above two dimensions, the roles
of controlling scale transformations and the short-distance singularity
are related to two different spin structures. These two
structures do not talk to each other since Lorentz invariance
acts separately on each one. This is the reason why we could
not find an independent characterization of our $c$-charge.

The previous problem can be made quantitative by checking
of $\Delta c$ in a chain of RG flows. For three fixed points
(fig.  1), this reads
\eqn\additi{
(\Delta c)_{1\to 2} + (\Delta c)_{2 \to 3} =
(\Delta c)_{1 \to 3}\quad.}
Additivity of the $c$-charge also amounts to integrability
of the system of beta-functions.
Equation \additi~ holds if $c$ at the theory 2 has the same value
independently of whether we approach this theory from 1 or from 3.
This is not obvious, because $(\Delta c)_{i\to j}$ is computed in the
$i$-th theory, so that we are comparing calculations in two a priori
different off-critical theories.

Nevertheless, suppose that the three fixed points lie in a region of the
space of theories parametrized by smooth coordinates. Indeed, this is
the case of perturbative calculations (by definition).
$\Delta c$ is a polynomial in the renormalized couplings, thus
the limit from off-criticality exists, or equivalently, trajectories can
be deformed at will to prove additivity.
Therefore, we can say that the $c$-theorem is proven for all kinds
of perturbation expansions, namely $\Delta c \ll 1$
and polynomial in the couplings.
Two examples of this kind will be discussed in the next section.

The drawback to the above comments is that the space of theories has
singularities and is probably not a manifold.
For example, in fig. 2 we imagine  comparing $\Delta c$ for two
non-perturbative flows into the massive phase ($\Delta c\sim 1$).
The trivial theory ($c=0$) can appear in several points of the coupling
space, thus the two paths can be topologically inequivalent.
The additivity property,
$(\Delta c)_{1\to 2} = (\Delta c)_{1\to 3}$, is not at all trivial
in this case. This kind of situation will be discussed in sect. 4.
\vfill\eject


\def\ps{\phi_s}
\def\po{\phi_\perp}
\newsec{Perturbative Flows in Scalar Theories for $2<d<4$}

In this section, we provide examples of perturbative RG patterns.
Two or more fixed points appear in a small region of
coupling space, and the $d$-dimensional $c$-function varies
infinitesimally while flowing among them.
The validity of the $c$-theorem is verified by using
various versions of the $\epsilon$-expansion.

First we study the Landau-Ginsburg-Wilson
RG pattern of the $r$-th multicritical points in the
$\lambda\ph^{2r}$ theory.
The comparison of the $c$-charges of the $(r)$ and $(r-1)$ theories suggests
that the known multicritical RG flows in two dimensions extend smoothly
to higher dimensions.
Next, the multicomponent $\lambda\ph^4$ theory is
considered, and eq. \additi~ for additivity of the $c$-charge is verified.
\bigskip
\newpar{MULTICRITICAL POINTS AND LANDAU-GINSBURG THEORY}

The Landau-Ginsburg action
\eqn\landaug{
S_{LG}= \int d^dx \left({1\over 2} (\partial_\mu\ph)^2
-\sum_{k=1}^r {\lambda_{k,0}\over 2k!} \ph^{2k}\right)\quad.}
describes the qualitative features of generic $r$-multicritical
points with parity symmetry (only), which appear for $2\le d<4$ \wilson.
At the multicritical point
$\{\lambda_{k,0}\}=\{0,0,...,\lambda_{r,0}\}$,
$r$ minima of the potential merge, which correspond to $r$
coexisting phases of the theory.
The flow to lower $r'$-critical points, $r'<r$, is described
by switching on the relevant perturbation $\lambda_{r',0}\ph^{2r'}$.
At the end of the flow, the higher powers $\ph^{2k}, \, r' < k \leq r$
become irrelevant fields and can be neglected in the action.

In two dimensions,  the minimal conformal theories \bpz~
with central charge \hfill\break\hbox{ $c(r)=1-{6\over r(r-1)}<1$}
are the exact renormalization of the above Landau-Ginsburg
actions \zamlg. The renormalized fields $\ph^{2k}$, $k<r$, are the
primary conformal fields which appear in the
first two diagonals of the Kac table.
Off-criticality, this picture has been confirmed along the flow between
the $(r)$ and the $(r-1)$ models, driven by the least relevant field
$\ph^{2r-2}$, for $r\gg 1$ \zamcth\ludwig\footnote*{As recalled in example
1 of the Appendix.}.
The dimension of this field is $2-\eps$,
$\eps\sim 1/r \ll 1$, thus it is the typical perturbation
situation encountered in the $\eps$-expansion. The IR fixed point,
$(r-1)$, appears infinitesimally close to the UV one, (r),  in
coupling space.  The invariant definition of distance is given by the
Zamolodchikov metric, eq. \zcth, or, equivalently, by the change of the
central charge, $\Delta c=c_r -c_{r-1}\sim O(1/r^3)$.

The $c$-theorem is a nice complement to the Landau-Ginsburg picture.
Higher multicritical points have higher values of $c$, thus flowing
downhill corresponds to going to lower multicritical points. Vafa has
further developed this picture \vafa.
The $c$-function can be considered as the
height function in the space of theories $Q$, so that Morse theory can
be applied and the Poincar\'e polynomial gives some information on the
holonomy of $Q$. In short, the qualitative Landau-Ginsburg description
together with the $c$-theorem give some grasp of the topology
of this space of theories.

Following Wilson, the Landau-Ginsburg picture holds
for all dimensions up to the upper critical dimension $d_c$
\eqn\upcritical{
2\leq d \leq d_c\of{r} \equiv 2 +{2\over r-1}\quad,}
the dimension for which $\ph^{2r}$ becomes marginal, {\sl i.e.} the
$r$-th multicritical point merges with the Gaussian one.
A natural question to
ask is whether our candidate for a $c$-theorem extends above two
dimensions as well. The higher multicritical points should continue to
have larger values of $c$. In such a case, the space of multicritical
points and their flows would have an analytical continuation in
dimensions.

\newpar{THE $C$-CHARGE OF THE $\lambda \ph^{2r}$ THEORY }

Above two dimensions, we compute our candidate $c$-charge $c_r$
of the r-th multicritical point by applying
the sum rule eq. \dsumrule~ to the flow from
the Gaussian theory, $c_r=1-\Delta c$ (fig. 3).
We use the $\eps$-expansion at dimension
\eqn\ddc{
d=d_c(r) -{\eps_r\over r-1}\quad,\quad 0<\eps_r \ll 1\quad,}
where $\eps_r={\rm dim}(\lambda_{r,0})$ is the small parameter.

Let us start with the familiar
example of $\lambda \ph^4$ theory, for $d=4-\eps$, and derive the
first order term of the $\eps$-expansion.
The action is
\eqn\actphifour{
S = \int d^d x \left[ \ha \of{ \partial_\mu \ph_0}^2
- {\lambda_0 \over 4!} \ph_0^4 \right],}
where $\lambda_0$ and $\ph_0$ are the bare coupling constant and field
respectively.
At $d=4-\eps$ the dimension of the coupling constant is $dim(\lambda_0)=
\eps$, {\sl i.e.} the field $\ph_0^4$ is slightly relevant and it
produces a flow from the Gaussian fixed point $\lambda_0=g=0$
($c=1$, by definition) to
the Wilson fixed point at $g=g^*\sim \eps$, where $g$ is the
renormalized coupling constant.

The trace of the energy momentum tensor is
\eqn\thetaphifour{
\Theta = -\eps {\lambda_0 \over 4!} V \ph^4.}
{}From the computation of the two leading Feynman diagrams, we can extract
\eqn\imaginarypart{
Im \left.\langle \Theta(p)\Theta(-p)\rangle\right\vert_{p^2=-\mu^2} = \eps^2
{(\lambda_0S)^2V^2S\pi\over 3\cdot 128} \left( {1\over 6} \mu^{4-3\eps}
+ {\lambda_0 S\over \eps} \mu^{4-4\eps}\right)\quad.}
Upon insertion of this imaginary part in eq. \dsumrule~ and integration,
one obtains
\eqn\cfunction{
c(\lambda_0 \kappa^{-\eps}) = 1- a \left(
\lambda_0S\kappa^{-\eps}\right)^2 \left( \eps +
(\lambda_0S\kappa^{-\eps}) b\right)\quad,}
\eqn\cost{
a =  {5\over 48} VS,\qquad
b = 4, \qquad
S \equiv {V\over(2\pi)^d} = {2\over (4\pi)^{d\over 2} \Gamma({d\over 2})}.}
This flowing $c$-function depends on the bare coupling and the IR cut-off
$\kappa$, which appears in the intermediate steps of calculations of
massless perturbations, as usual \ludwig.
Since $c$ has no anomalous dimension,
its renormalization is simply achieved by replacing
$\lambda_0\kappa^{-\eps}$ with the renormalized coupling $g$. This is a
change of coordinates in coupling space which removes the unphysical
singularity in the Zamolodchikov metric $G(\lambda_0)$ at the IR fixed
point \lassig. Because of eq. \zcth~, all the information needed to find such a
transformation is contained in the $c$-function itself, and we need not
carry out the renormalization of fields. This produces economical
formulae for the flow.
For one-coupling flows, eq. \zcth~ gives
\eqn\partiallambda{
{\partial\over \partial \lambda_0}c(\lambda_0) =
G(\lambda_0)\beta(\lambda_0)\quad,}
where the beta-function in terms of the bare coupling constant is
$\beta(\lambda_0)=-\eps \lambda_0$, by eq. \thetaphifour.
The relation between bare to renormalized couplings can, then, be cast
into an elegant geometrical condition -- the invariant distance in
coupling space remains the same whatever coordinate system is chosen,
\eqn\invariantdistance{
ds^2 = G(\lambda_0) d\lambda_0^2 = G(g) dg^2\quad.}
The renormalized coupling is defined by requiring that $G(g)=2a$, where the
specific value chosen for the constant is of later convenience. Then
$g(\lambda_0)$ is obtained by integrating eq. \invariantdistance.
The final form for the $c$-function reads
\eqn\cfinal{
c(g)=1-ag^2\left(\eps + {b\over 4}g\right) = 1-{5\over
3\cdot 64} g^2 (\eps+g)\quad.}
Its derivative is proportional to the beta-function,
\eqn\betaphifour{
\beta \of{g} =-\eps g - {3\over 8} b g^2 =
 - \eps g - {3 \over 2} g^2 ,}
which agrees with standard derivations (see {\sl e.g.}\zinn).
In agreement with the general discussion of eq. \zcth \cfl,
we have obtained a monotonic decreasing $c$-function, which is stationary at
the fixed points  $g=0$ and $g=g^*= -{2 \over 3} \eps$. These results
are an explicit illustration of Zamolodchikov's ideas in more than two
dimensions \zamcth.

The value of the $c$-function  at the Wilson $(2)$-critical point is
\eqn\deltacphifour{
c_{2} = c(g^*) =
1 -  {5 \over 16 \cdot 81} \eps^3 \quad, d=4-\eps\quad.}

\medskip
We  can generalize this analysis to the flow between the Gaussian and
the $(r)$-critical point. The action is
\eqn\actphitwor{
S = \int d^d x  \left[ \ha \of{ \partial_\mu \ph}^2
- {\lambda_{r,0} \over 2r!} \ph^{2r} \right],}
where the dimension now is slightly below $d_c(r)$, eqs.\upcritical,\ddc.
The trace of the stress tensor reads
\eqn\thetaphitwor{
\Theta =-\eps_r {\lambda_{r,0} \over 2r!} V \ph^{2r}.}
Again the computation of the two-point correlator to leading perturbative
 order involves two
Feynman diagrams  which have
the same singularity structure as the $r=2$ case,
implying again stationarity of $c(g)$. The only numerical changes are
\eqn\aandb{
\eqalign{ a_r &= 2^{d-3} \of{d + 1} V \left(
{S \left[ r \right] \left[ 1 \right] \over 2} \right)^{2r-3}
{\left[ 1\right]^3 \left[ r\right]^2\over \left[ 2r-1 \right] 2r! } \cr
b_r &= {4 \over 3}  \left(
{S \left[ r \right] \left[ 1\right] \over 2} \right)^{r-2}
{\left[ 1 \right]^2  2r! \over (r!)^3} \cr}}
where the notation
$\left[ \alpha \right] = \Gamma \of{ \alpha \over r-1}$
has been introduced\footnote*{
The corresponding beta-function agrees with ref. \itzykson~ by
rescaling $g\rightarrow g/S$.}.
Again we find a non-trivial fixed point at
$g_r=g^*_r \equiv - 8 \eps_r / 3 b_r$, with $c$-charge
\eqn\deltacphitwor{
c_r= c(g^*_r) = 1 -  {3r-1 \over 3r}
\left({r!^2 \over 2r!}\eps_r \right)^3 \quad,\qquad
d={2r - \eps_r \over r-1}, \quad 0<\eps_r \ll 1.}

Notice that the $\epsilon_r$-expansion to first order
is not good enough for reproducing the known values of the charge
in two dimensions, especially for large $r$.
We got \hbox{$c_r \sim 1- O(2^{-6r})$} instead of $c_r \sim 1- O(1/r^2)$.
Actually, the first few terms of this asymptotic expansion
give an accurate result for
$0\le \epsilon_r < O(A^{-r+1})$, where $A$ is a positive constant,
so that we cannot use it for two dimensions $(\epsilon_r=2)$\footnote{**}{
This bound can be obtained from the fact that the $k$-th term in the
expansion grows like $(k!)^{r-1}$ \lipatov.}.

\newpar{THE HEIGHT OF THE $(r)$-THEORY VERSUS THE $(r-1)$ ONE}

We have now the elements to compare the $c$-number or ``height" of two
neighbour multicritical points. From the Landau-Ginsburg picture and the
two-dimensional $c$-theorem we expect a flow from the $(r)$-theory to
the $(r-1)$ one (see fig. 3).
If the $c$-theorem holds in any dimension we expect
$c_r > c_{r-1}$. At dimension
\eqn\moredim{
d= d_c(r) -{\eps_r\over r-1} = d_c(r-1) - {1\over r-2} \left(
{2\over r-1} + {r-2\over r-1} \eps_r\right) \qquad 0<\eps_r \ll 1.}
both critical points are present (fig. 3) and we find
\eqn\crossing{
c_r - c_{r-1} =
\left( {r!^2\over 2r!}\right)^3
\left[ {\left(1-{4\over 3r}\right) \left(1-{1\over 2r}\right)^3
        \over \left(1-{1\over r} \right) }
\left({8\over r-1} + 4 {r-2 \over r-1} \eps_r \right)^3 -
\left( 1-{1\over 3r}\right) \eps_r^3\right] \, >0\quad.}
This bound is satisfied for all $\eps_r$ within the
range of validity of the $\eps$-expansion at first order.
As shown by fig. 3, both multicritical points exist
for $\eps_r$ small at will, and $\eps_{r-1}\sim O(1/r^2)$ at least,
the latter going outside the range of accuracy for large $r$.
Therefore our result in eq. \crossing~ is probably numerically good
for small $r=2,3$, but only heuristic for large $r$.
Nevertheless, the comparison of the two charges at the same
perturbative order is certainly better than the absolute value
of each one. Eq. \crossing~ shows that both $c$-charges have
exponentially small corrections, but differ in an algebraic factor.

Moreover, $c_r >c_{r-1} $ necessarely holds for $\eps_r\to 0$.
Indeed, the nucleation of multicritical points ordered in dimension
(see fig. 3), the fact that $c=1$ for the free theory in any dimension and
its monotonicity property imply $c_{r-1} <1$ and $c_r\sim 1$ in this limit.

Let us finally quote Felder's investigation of this RG pattern
in more than two dimensions
\felder. Working in a different perturbative approach, he was also able
to build a monotonic function along the RG flows from his system of
beta-functions.

In conclusion, these are good indications that the topology of the
space of multicritical scalar theories extends smoothly above two dimensions.

\newpar{THE MULTICOMPONENT $\ph^4$ THEORY}

Another well-known RG flow pattern is provided by the multicomponent
$\ph^4$ model in $4-\eps$ dimensions \brezin,
\eqn\multicomponent{
S=\int d^dx\left( {1\over 2} (\partial_\mu \ph_i)^2
-{1\over 4!} g_{ijkl} \ph_i\ph_j\ph_k\ph_l\right)\quad,}
where the sum over $N$ components $\ph_i, i=1,...,N$ is implicit. In
this theory, Wallace and Zia \wz~ first showed that the RG trajectories are
gradient flows to three-loop order in the
$\eps$-expansion,
thus ensuring that the flow is
driven to the IR fixed points -- a non-trivial fact when we look at the
multi-loop $\beta$-function!

These results can easily be framed into the $c$-theorem philosophy. A
simple modification of our previous Feynman rules leads to
\eqn\cmulti{
c\of{g_{ijkl}}= 1-{5\over 64\cdot 3}\left( \eps g_{ijkl}
g_{ijkl} + g_{ijkl} g_{ijrs} g_{klrs}\right)}
and
\eqn\bmulti{
\beta_{ijkl}(g_{mnrs}) = {1\over 2a} {\partial\over \partial
g_{ijkl}}c = -\eps g_{ijkl} -{3\over 2} g_{ijrs} g_{klrs}\quad,}
which necessarily agree with the Wallace and Zia result to one loop. To
higher orders, our $c$-function would correspond to a definite choice of
the free parameters in their gradient function $\Phi$,
$c\sim 1-{\rm const.} \Phi$.

The RG flows given by eq. \bmulti~ correspond to trajectories leaving the
Gaussian point in different directions and reaching IR fixed points. One
of them is stable and, therefore, displays a minimum of $c$. The other
ones are unstable, the stability changing with $N$.
Let us recall the results of ref. \brezin. A two-dimensional subspace of
the coupling space is given by the $O(N)$-symmetric
perturbation $\left( \sum_i\ph_i^2\right)^2$ leading to the
$O(N)$-symmetric Wilson point, and the hypercubic symmetric one
$\sum_i\ph_i^4$,
leading to $N$ decoupled Ising models. This pattern describes the
breaking of $O(N)$ symmetry to the hypercubic one in lattice
ferromagnets.
After computing the location of the $O(N)$- symmetric and
hypercubic fixed points, it is convenient
to introduce rescaled variables $x$ and $y$,
\eqn\xandy{
g_{ijkl}\ph^i\ph^j\ph^k\ph^l= -x{6\eps\over
N+8}\left(\sum_i\ph_i^2\right)^2 - y {2\eps\over 3} \sum_i\ph_i^4\quad,}
such that the $O(N)$ symmetric point is at $(x,y)=(1,0)$ and the
hypercubic symmetric is at $(x,y)=(0,1)$ (fig. 4).

The $c$-function in this parametrization  is
\eqn\cxy{
c \of{x,y} = 1 - {5 \over 16 \cdot
3} \eps^3
\left[ {N \of{N+2} \over \of{N+8}^2} \of{3x^2 - 2x^3} +
{2N \over N+8} xy \of{1 - x -y} + {N \over 27} \of{3y^2 - 2y^3} \right].
}
Inspection of the extrema of $c$ ( which correspond to
$\beta_x=0,\beta_y=0$) shows that there is a fourth fixed point at
$(x^*,y^*)=\left({N+8\over 3N},{N-4\over N}\right)$. The stability of
these points is also deduced from $c$. For any value of $N$, there
are two unstable points besides the Gaussian one and a stable fixed
point. For $N<4$, the $O(N)$ symmetric point is stable; for $N>4$,
$(x^*,y^*)$ is stable. An example of the shape of $c(x,y)$ for $N=8$ is
shown in the level plot fig. 5.

\newpar{ADDITIVITY OF THE $C$-CHARGE}

Let us now verify the additivity property of the $d$-dimensional
charge for compositions of flows among three fixed points, eq. \additi,
\eqn\additwo{
(\Delta c)_{1\to 2} + (\Delta c)_{2 \to 3} =
(\Delta c)_{1 \to 3}\quad.}
In the previous RG pattern, let us consider two possible chains:

i) For $N > 4$, we can reach the stable IR fixed point $(x^*,y^*)$
in two different ways, fig.  4a,
\eqn\cha{
(0,0) {\longrightarrow} (1,0){\longrightarrow} (x^*,y^*) \qquad {\rm or} \qquad
(0,0) {\longrightarrow} (x^*,y^*)}
corresponding to the l.h.s. and r.h.s. of eq. \additwo~.

ii) For $N>10$, we can also compare, fig. 4b,
\eqn\chb{
(0,0) {\longrightarrow} (1,0){\longrightarrow} (0,1) \qquad {\rm versus} \qquad
(0,0) {\longrightarrow} (0,1)}

Remember that each $(\Delta c)_{i\to j}$ in eq. \additwo~ has to be computed
using
the sum rule eq. \dsumrule~ applied to the corresponding off-critical theory,
having the $i$-th fixed point as UV limit.
Within the coordinates $(x,y)$ considered so far, the UV fixed point was
the free Gaussian theory $(0,0)$.
Thus, without extra work we
can particularize eq. \cxy~ for the flows
\eqn\deltacxy{ \eqalign{
 c(0,0) - c(1,0) &= {5 \over 48} { N \of{N+2} \over \of{N+8}^2}\eps^3 \cr
 c(0,0) - c(0,1) &= {5 \over 48} { N \over 27}\eps^3 \cr
 c(0,0) - c(x^*,y^*) &= {5 \over 48}{ (N+2)(N-1)\over 27 N} \eps^3. \cr}}

The other flows appearing in eqs.\cha,\chb~ should be described
in another coordinate patch, having $(1,0)$ as UV fixed point.
A more advanced perturbative
technique is needed, because the starting theory is
not free. Nevertheless, we can make use of the fact that the $(1,0)$
fixed point is conformal invariant (the trace of the stress
tensor vanishes) and use the ``conformal perturbation theory'' (see
 the Appendix and ref.\mavro).
This technique generalizes the one used for flowing off conformal theories
in two dimensions  \zamcth\cl.
Conformal invariance in higher dimensions fixes
the form of 2- and 3-point functions, up to some coefficients, the
conformal dimensions of fields and the structure constants.
The first order perturbative expansion only requires these ingredients, thus
it can be given for any flow in any dimension.

Let us first build a suitable basis of
operators around the $(0,0)$ fixed point.
We choose the orthonormal basis
\eqn\opbasis{
\ps^{(0,0)} \equiv A\,\left(\sum_i \ph_i^2\right)^2 \qquad,\qquad
\po^{(0,0)} \equiv B\, \left(\sum_i \ph_i^4
   -{3\over N+2} \left( \sum_i \ph_i^2\right)^2 \right),}
$$A = \sqrt{3 \over N(N+2)} \qquad,\qquad
     B = \sqrt{N+2\over N(N-1)}\quad, $$
which we call ``symmetric" and ``orthogonal" respectively. It is
easy to compute the three-point functions made out of these two
operators, and obtain the following structure constants
\eqn\threepf{\eqalign{
& C_{sss}^{(0,0)} = {A(N+8)\over 9}\, C \qquad,\qquad
  C_{ss\perp}^{(0,0)} = 0 \,,\cr
& C_{s\perp\perp}^{(0,0)} = {A(N-2)\over N+2}\, C \qquad,\qquad
  C_{\perp\perp\perp}^{(0,0)} ={ B(N-2)\over N+2}\, C\quad,\cr
}}
where $C$ is the $N$-independent part of the structure constant.
Note that, to the order in $\epsilon$ we are working, the
structure constants remain the same at the (1,0) fixed point,
{\sl i.e.}
\eqn\identppp{
C_{ijk}^{(0,0)} \, =\, C_{ijk}^{(1,0)} \qquad i,j,k=s,\perp}
and, therefore, we can omit these superindices from now on.

Using the orthonormal basis we have just presented, we use the
$\phi_s$ field as a perturbation off the Gaussian fixed point. This
perturbation leads us to the (1,0) point.
In general, conformal perturbations are subject to RG
mixing of operators. Nevertheless,
the flow between (0,0) and (1,0) is free of this intricacy because
$\langle \ps\po\rangle^{(0,0)}=0$ and $C_{s\perp s}=0 $, {\sl i.e.}
$\ps$ and $\po$ stay orthogonal all the way along the RG flow\footnote*{
Actually, in this case the general system of beta-functions
given in the Appendix simplifies to a single equation.}.

It is easy to compute the scaling dimensions of the symmetric and
the orthogonal operators which get renormalized,
\eqn\andim{ \eqalign{&
\Delta_s^{(1,0)}  = \Delta_s^{(0,0)} + 2 \left( d-\Delta_s^{(0,0)}\right)
 = d+\epsilon \qquad\equiv d -y_s \cr
& \Delta_\perp^{(1,0)} =  \Delta_\perp^{(0,0)} + 2 \left( d-\Delta_s^{(0,0)}
\right){C_{\perp\perp s}\over C_{sss}} =
d- {N-4\over N+8}\eps \equiv d-y_\perp \quad.\cr}}
The above
equations tell us that the symmetric operator becomes irrelevant
at the (1,0) fixed point whereas the orthogonal one is relevant
for $N > 4$, in agreement with the discussion of the phase diagram.

Having established the new basis of fields at (1,0), we can now compute
the flows off this point driven by $\po^{(1,0)}$ for $N>4$.
In this case,
though $\langle \ps \po\rangle^{(1,0)}=0$, we note that
$C_{p\perp\perp}\not= 0$, which implies that $\ps^{(1,0)}$ and
$\po^{(1,0)}$ do mix along the new flow. Thus, we are forced to consider
a general two-parameter deformation of the (1,0) fixed point, namely
$$ g_\perp \po^{(1,0)}+ g_s \ps^{(1,0)} \quad. $$
The system of beta-functions one obtains in conformal perturbation
theory is (see the Appendix)
\eqn\betas{ \eqalign{
& \beta_\perp =-y_\perp g_\perp - \left(
C_{\perp\perp\perp} g_\perp^2 + 2 C_{\perp\perp s} g_\perp g_s\right)
\cr
&\beta_s = -y_s g_s -\left(C_{s\perp\perp} g_\perp^2 + C_{sss} g_s^2
\right)\quad.\cr}}
The above set of equations correspond to the new $c$-function for
flows off the $(1,0)$ fixed point, call it $\tilde c$,
\eqn\cfu{
\tilde c(g_\perp,g_s) = c(1,0)
- y_\perp { g_\perp^2\over 2}- y_s {g_s^2\over 2}
-\left( C_{\perp\perp\perp} {g_\perp^3\over 3} +
C_{\perp\perp s} g_\perp^2 g_s + C_{sss} {g_s^3\over 3}\right)}
It turns out that the system of beta-functions has two solutions,
\eqn\betasol{
 (g_\perp^*,g_s^*) =\cases{ - {\epsilon\over C} \left(
{N-4\over B N}, {A(N-4)^2\over (N+8)}\right) \cr
 - {\epsilon\over C} \left(
{1\over B}, {2AN(N-1)\over (N+8)} \right)\cr}}

Finally, we can substitute back
the solutions of the $\beta_s,\beta_\perp=0$ system
in the $\tilde c$-function and get,  for the two solutions,
\eqn\chainc{
\Delta \tilde c= \cases{
{5\over 48} {(N+2)(N-4)^3\over 27 N(N+8)} \epsilon^3
\quad \,\, = c(1,0) - c(x^*,y^*) \cr
{5\over 48}{N(N-1)(N-10)\over (N+8)^2}\epsilon^3
\quad = c(1,0) - c(0,1) \cr} }
The comparison of this result to eq. \cxy~ shows that
the two solutions correspond to the two flows in
eqs.\cha,\chb~  and, therefore,
the additivity property of our $c$-charge,
eq. \additwo, does hold.

We are now confident of the limiting procedure eq. \clim~
which defines the
$c$-charge at fixed points from the spectral function $c(\mu,\Lambda)$
away from criticality $(\Lambda\not= 0$).
In a perturbative domain, it produces consistent results
for inequivalent coupling coordinates, so that the $c$-charge is
indeed a universal quantity attached to each fixed point.

These results, though expected,
have tought us that,
 in general, we have to
 allow for irrelevant fields in the expansion of $\Theta$
in eq. \trace~, if they mix with
the relevant one driving the flow.

\vfill\eject


\def\i{\sf i}
\newsec{The $O\of{N}$ sigma-model in the Large $N$ Expansion}
In the previous section we considered RG flows between infinitesimally
close fixed points, $\Delta c \ll 1$. The large $N$ expansion allows,
instead, to describe RG flows which run over a large
distance in coupling space, {\sl e.g.} $\Delta c \sim N$,
thus non-perturbative with respect to the coupling. Actually, the saddle
point method amounts to the resummation of an infinite set of
diagrams of conventional perturbation theory, and it leads to
beta and $c$ functions which are non-analytic in the coupling
(the mass for the sigma-model).
Further corrections in the $1 \over N$ expansion are
similarly non-analytic for what concerns the $c$-theorem and the sum rule.
In the following, we study the flow in the symmetric
phase of the $O \of{N}$ sigma-model, for large $N$
and $2\leq d\leq 4$.

The $O\of{N}$ symmetric non-linear sigma-model is defined by an
action which contains  $N$ fields $\ph_i$ and a Lagrange multiplier
$\alpha_0$
\eqn\zon{ \eqalign{
Z &= \int \CD \alpha_0\,\CD \ph^i \e{-S}\cr
S &= \int d^d x \ha \left[ \partial_\mu \ph^i \partial_\mu \ph^i
+ \alpha_0 \of{ \ph^i \ph^i - {N \over g_0^2}} \right],}}
where the fields and the coupling are conveniently rescaled for the
large
$N$ expansion. The integration over $\ph_i$ produces the effective
action
\eqn\seff{
S_{eff} = {N\over 2} \ln \det \null_\Lambda \of{ -\del^2 +
\alpha_0(x)} - {N \over 2 g_0^2} \int d^d x \alpha_0\of{x}.}
which, for large $N$, can be estimated using the saddle point
approximation. The saddle point equation is
\eqn\saddleeq{
{1 \over g_0^2} = \int^\Lambda {d^d p
\over \of{2 \pi}^d} {1 \over p^2 + m^2},\qquad
m^2=\vev{\alpha_0}_{s.p.},}
where $\Lambda$ is the cut-off, and $m^2$ is the translational
invariant value of the field $\alpha_0$ at the saddle point. The $1/N$
expansion is obtained by setting
$\alpha_0(x) = m^2 + {\alpha (x)\over\sqrt{N}}$
and expanding $S_{eff}$ around the saddle point. The
Feynman diagrams and the basic properties of the theory are discussed
in ref. \on. Investigations of the $c$-theorem for this theory were initiated
in ref. \cfl.

For $2 \leq d \leq 4$, the saddle point defines the physical mass $m$ for
$\ph$, in terms of the bare coupling $g_0^2$, while the higher order
corrections to the saddle point are weighted with the coupling
$1 \over N$.
{}From the heat kernel regularization of the determinant in eq. \seff~
we obtain
\eqn\computedseq{ {1 \over g_0^2} = {2 \over \of{4 \pi}^{d \over 2}}
\left[ {\Lambda^{d-2} \over d-2} - {m^{d-2} \over d-2} \Gamma \of{
2-{d\over 2}} - {\Lambda^{d-4} m^2 \over d-4} + \sum_{k=2}^\infty {1
\over k!} { \Lambda^{d-2k-2} \of{-m^2}^k \over d-2k-2} \right].}
In particular, for $d=2$,
\eqn\seqtwod{ {1 \over g_0^2} = {1 \over 4 \pi} \ln \of{{
\Lambda^2 + m^2 \over m^2}}.}
The critical point $g_{0,cr}^2$ is obtained for $m=0$
\eqn\critcoupl{ g_{0,cr}^2 = 0 \quad \of{d=2}, \qquad {1 \over
g^2_{0,cr}} = {2 \over \of{4 \pi}^{d \over d}} { \Lambda^{d-2} \over d-2}
\quad \of{d>2},}
and the massive phase corresponds to $g_0^2 > g_{0,cr}^2$. On the other
hand, for $g_0^2 < g_{0,cr}^2$, there is no solution to the $O
\of{N}$-symmetric saddle point equation. There are non-symmetric
saddle points, obtained by integration only $N-1$ fields in eq. \zon,
giving $\vev{\ph} \not= 0$ \on.
In short, the phase diagram of the sigma-model for $2 \leq d \leq
4$ contains the $O\of{N}$-symmetric phase $g_0^2 > g_{0,cr}^2$, with
$\vev{\alpha_0}=m^2$, $\vev{\ph}=0$, and (for $d>2$), the spontaneously
broken one $g_0^2 < g_{0,cr}^2$, with $\vev{\alpha_0} =0$ and
$\vev{\ph} \not= 0$.
\bigskip

4-1. THE STRESS TENSOR AND THE SPECTRAL DENSITY

Let us consider the RG flow in the symmetric phase which leaves
the critical point and reaches  the trivial fixed point $m= \infty$.
We want to compute the associated sum rule eq. \dsumrule,
to leading order in $1 \over N$.
We start by finding
the expression of the trace of the stress tensor for this perturbation
in terms of the $\alpha$ field. This is obtained by putting
the theory on a curved space-time background and taking a Weyl
variation with respect to  the metric of both the effective action
and the saddle-point equation \cfl.
Furthermore, for $2<d<4$ we let the cut-off go to infinity after
one subtraction of the coupling. We obtain\footnote*{
Note that $\Theta$ vanishes classically for $d=2$,
but not at the quantum level (saddle point).
We renormalize the theory by going to $d=2+\eps$, and
find a non-vanishing beta-function to leading order in $1/N$.},
\eqn\thetaon{\eqalign{
\langle \Theta (x)\rangle &= \beta (m)\, \langle \alpha (x)\rangle_{s.p.}=0\cr
\langle \Theta (x)\Theta(0)\rangle &=
\left(\beta (m)\right)^2 \,\langle \alpha(x)\alpha(0)\rangle_{s.p.}\quad,
x^2 \neq 0 \quad, \cr}}
where
\eqn\betaon{\beta(m)=
m^{d-2}\,\sqrt{N}\, {\Gamma\left( 2-{d\over 2}\right) \over
  2^{d-1} \Gamma\left({d\over 2}\right)} \quad.}

The leading contribution to the $\alpha$ propagator is obtained by
expanding $S_{eff}$ to second order in $\alpha$
\eqn\alphaalpha{ \vev{ \alpha \of{p} \alpha \of{-p} }_{s.p.}
= - {2 \over B \of{p}},}
where $B\of{p}$ is the well-known bubble diagram
\eqn\blob{ \eqalign{
B \of{p} &= \int {d^d q \over \of{2 \pi}^d} {1
\over \of{q^2 + m^2} \of{\of{p-q}^2 + m^2}} = \int_{4m^2}^\infty
{d\mu^2 \over \pi} { \impart B\of{p^2 = - \mu^2} \over p^2 + \mu^2} \cr
         &= {m^{d-4} \Gamma \of{{4-d \over 2}} \over 2^d \pi^{d\over 2}}
F\of{1, {4-d \over 2}; {3 \over 2}; -{p^2 \over 4m^2}} \cr
\impart B\of{\mu^2} &= {1 \over \Gamma \of{{d-1 \over 2}} 2^d \pi^{ d-3
\over 2}} {1 \over \mu} \of{ {\mu^2 \over 4} - m^2}^{d-3\over 2} \theta
\of{\mu^2 -4m^2} \cr}}
where $\theta \of{x}$ is the step function and $F$ the hypergeometric
function.
\bigskip
\bigskip

4-2. THE SUM RULE IN TWO DIMENSIONS
\bigskip

In two dimensions, the critical point is $g_0=0$, and there is
only the symmetric phase, in agreement with  Coleman's theorem on the
absence of spontaneous symmetry breaking.
Around the critical point, the theory contains $N-1$ weakly interacting
bosons, asymptotically free \zinn, thus we can assign the value
$c_{UV} = N-1$.
The symmetric phase is massive and the IR Hilbert space contains $N$
interacting massive scalar particles with purely elastic scattering,
thus $c_{IR} = 0$.
Therefore, the $c$-theorem sum rule is expected to give $\Delta c= N$
for the flow in the symmetric phase, to leading order in $1/N$.

In order to compute it, we need the spectral function $c\of{\mu}
\propto \impart \vev{\Theta \Theta}$,
given by eqs.\thetaon,\alphaalpha.
The imaginary part of the $\alpha$-propagator is
$\impart \of{ -2 \over B} = 2 \impart \of{B} /\abs{B}^2$,
where $B\of{p}$ can be expressed in terms of logarithms for $d=2$.
Putting all together, we obtain
\eqn\specdens{ \eqalign{ c \of{\mu}
&= {6 \over \pi^2 \mu^3}
\at{Im \vev{\Theta \of{p} \Theta \of{-p}}}{p^2+\mu^2=0} \cr
&= {6N \over \mu} \sqrt{1 - {4m^2 \over \mu^2}}
\left[ \pi^2 + \log^2 \left(
1 + \sqrt{1 - {4m^2 \over \mu^2}} \over
1 - \sqrt{1 - {4m^2 \over \mu^2}} \right)\right]^{-1} \,
\theta \of{\mu^2 -4m^2}. \cr}}
The sum rule is
\eqn\intdeltac{
\Delta c = \int_0^\infty d\mu c\of{\mu}
= 3N \int_{-1}^1 {dz \over 1-z^2} {z^2 \over \pi^2 + \ln^2 \of{{1+z
\over 1-z}}}. }
After a change of variable, the integral can be computed in
the complex plane, giving the expected result
\eqn\dctwo{
\Delta c = N, \qquad\qquad (d=2,\qquad  N\to\infty).}
This is a rather remarkable check of Zamolodchikov's theorem,
owing to the non-perturbative character of this flow.

In this calculation we had no problems with IR singularities,
which instead can appear in the perturbation expansion\footnote*{It
 has been suggested that the $c$-theorem could fail because of them \curci.}
in $g_0^2$ .
Actually, the large $N$ expansion is better
because it correctly reproduces the mass of the theory by resumming
infinite perturbative diagrams.
Let us however stress that IR singularities cannot, in general, appear in the
$c$-theorem, unless the theory itself is sick and plagued with them.
As remarked in sect. 2, our formulation of the $c$-theorem
makes use of the states of the Hilbert space,
which would be ill-defined in the presence of IR singularities.
\bigskip\bigskip

4-3. THE SUM RULE ABOVE TWO DIMENSIONS
\bigskip

The sum rule can also be computed for $2<d\leq4$ by using the previous
formulae.
Before that, let us find the value of the $c$-charge at the critical
point of the sigma-model.
Having no intrinsic means to compute it, we have to relate it to
the value of the Gaussian free theory, $c=N$ in our conventions.
We use a number of arguments known in the literature \zinn~ to
embed the sigma-model into its linear realization, the
$O(N)$-symmetric $\lambda\ph^4$ theory (fig. 6).
The latter theory has a renormalized mass parameter $m'$,
and the additional interaction strenght $\lambda$, while the unique
mass parameter of the non-linear sigma-model has to be thought as
$m=m(m',\lambda)$, at least in a region close to the critical point.
The flow in the sigma-model in the symmetric phase $m\to\infty$,
will correspond to the line drawn in the $(m',\lambda)$ plane of fig. 6
(possibly leaving the plane for large $m$).
For $2<d\leq 4$, the sigma-model critical point $m=0$ is believed to fall
into the universality class of the $O(N)$-symmetric Wilson point of
$\lambda\ph^4$, discussed in sect. 3.4, {\sl i.e.}
$\left(m'=0, \lambda=\lambda^* = O({d-4\over N})\right)$.
The other fixed point of $\lambda \ph^4$, the Gaussian one,
is close to it for large $N$, and the corresponding flow between them
was previously shown to give $\Delta c=O(1)$, which is subleading
for large $N$.
Therefore, we can assign $c=N$ to the critical point
of the sigma-model for large $N$.

Actually, we are testing the $c$-charge along a chain of flows as in
eq. \additi,
where the fixed points are: $(1)$ the Gaussian one, $(2)$ the critical
sigma-model, and $(3)$ the trivial theory $c=0$ (see fig. 6).
The last one actually sits in two different points of coupling space,
$(m'=\infty, \lambda=0)$ and $(m=\infty)$, thus we are in the
situation of non-perturbative flows of fig. 2.

Let us first discuss the sum rule in four dimensions, where
the Wilson point merges with the Gaussian one for any $N$.
At $d=4$ a new logarithmic singularity appears in the saddle point equation
eq\saddleeq.
According to the previous discussion (eqs.\thetaon), we regulate it
by using dimensional regularization. Thus we find,
\eqn\thetafourd{ \eqalign{
\Theta \of{x} & \sim {\sqrt{N} \over 4}
{m^{2-\eps} \over \eps} \alpha \of{x} \qquad, \qquad\qquad (d=4-\eps) \cr
B\of{p} &\sim {1 \over \eps} {m^{-\eps} \over 2 \pi^2} \left( 1 + O \of{
\eps {p^2 \over 4 m^2}} \right). \cr}}
The singularities $1 / \eps^2$ cancel in the spectral density
$c\of{\mu}$, which has a finite expression in any dimension, as it should.
Finally,
\eqn\deltacfourd{
\Delta c = N \int_0^\infty d\mu {c \of{\mu} \over
\mu^2} = N \int_{2m}^\infty d\mu\, 120 \, {m^4 \over \mu^5 }
\sqrt{ 1 -{4m^2\over \mu^2}} = N \quad,\qquad (d=4, N\to\infty) }
Indeed, $c\of{\mu}$ has the same form as in the massive perturbation
of the free four-dimensional theory, computed in ref. \cfl
(see eq. (3.31) therein).
Therefore, the sigma-model approaches the free theory all along
the massive flow, and the sum rule confirms the expectations on the
triviality of the model, to leading order in $1 \over N$.
Moreover, the test of addivity of the $c$-charge eq. \additi~
is trivial in this case.
\bigskip
\bigskip

As an example of the case $2<d<4$, we shall discuss $d=3$, where
eqs.\blob~ can be expressed in terms of elementary functions, and
the final integral of $c(\mu)/\mu$ computed numerically.
This was already done in our previous work \cfl~, eq. (6.47), and
we only quote the result,
\eqn\sumthree{\Delta c=N\, (0.5863....)\qquad,\qquad (d=3, \, N\to\infty)
}
Therefore, we do not find agreement with the expected result $\Delta c=N$.
A possible explanation of this failure is, of course, that our
candidate $c$-charge does not fulfil the addivity property, thus it is
not a universal quantity uniquely associated to the fixed points.
Indeed, in this case of non-pertubative flows, we do not have
arguments in support of additivity, nor  can we exclude
coordinate singularities in the space of theories of fig. 6.

Another explanation could be that the $c$-theorem, though correct,
is not easy to verify for non-perturbative flows.
In particular, we cannot exclude mixing with irrelevant fields
in the expansion of $\Theta=\beta^i\Phi_i$,
due to our partial understanding of the critical field theory
of the sigma-model.
Actually, the example of sect.  3.4 showed that irrelevant fields
can appear, when flowing off interacting critical theories.
The fact that the sum rule \sumthree~ has a value of $\Delta c$
lower than expected  may be an indication that we are missing contributions
to $\Theta$.
\vfill\eject

\newsec{Conclusion}

In this paper, we have put forward a candidate for a monotonically decreasing
function along RG trajectories in $d>2$, which is the analogue of the
Zamolodchikov $c$-function  in two dimensions.
The analysis of this $c$-function in the Ginsburg-Landau
models  using  epsilon expansion as well as conformal perturbation theory
displays all the nice properties expected from a $c$-theorem.
Our $c$-function behaves as a height function in a perturbative domain
of the space of theories.

The study of the sigma-models is  less clear. In two dimensions, the
theorem works as it should, proving harmless the fears concerning infrared
problems. In four dimensions, it confirms that the theory coalesces
with the free massive one and there are, again, no problems.
Nevertheless,  in three dimensions, our computation of ref. \cfl~
presents a result which is in disagreement with the theorem.
Different ways out are sketched in sect. 4, which deserve further
investigation.

There is an observation we want to emphasize. Physically
meaningful theories, like QCD, behave at short and long distances as
free theories\footnote*{In fact, there are fewer non-trivial scale
invariant theories in four dimensions, in contradistinction with
many non-Gaussian fixed points known to exist
 in less than four dimensions.}.
Remarkably enough,  it  was found in ref. \cfl~
 that the $c$-charges associated to the spin 0 and 2 spectral
densities are equal for free theories of spin 0 and 1/2. It is, then,
natural to conjecture that both  $c$-charges are equal in general, that is
$c^{(2)}=c^{(0)}$ for free theories. Since $c^{(2)}$ is well-defined at
fixed points,
a general proof stating this equality for free theories would be enough
to apply the $c$-theorem to the Standard Model and beyond.

As an example, it is easy to see that
 the long-distance Nambu-Goldstone realization of QCD in terms
of pions is in agreement with the conjectured $c$-theorem.
Adapting a previous example \cardycth, we
consider QCD with $N_f$ flavors and $N_c$ colours in four dimensions.
Using the fact that $c_{0}=1$ for particles with spin 0 and
$c_{1/2}=6$ for spin 1/2, we look at the balance between
the short- and long-distance realizations of
QCD
$$ N_f N_c \, c_{1/2} + (N_c^2-1) c_1 \qquad \geq \qquad (N_f^2-1) \, c_0 $$
for any value of $c_1$, provided asymptotic freedom holds,
{\sl i.e.} $N_f < {11\over 2}\, N_c$.
Note that we did not write the value of $c_1$, because
we cannot easily compute this number by a free massive perturbation,
as done in ref. \cfl~ for the spin 0 and 1/2 particles.
The massless limit of the massive spin 1, or Proca, particle
is not the massless spin-zero particle, because
the number of degrees of freedom changes.
Probably, one has to resort to the Higgs mechanism to give mass
to a gauge field in a correct way.

To conclude, we would like to mention some lines to progress.
The inclusion of additional symmetries ({\sl e.g.} current algebra)
remains to be done. Further analysis of the constraints
imposed by the present form of the $c$-theorem in
theories which go beyond the Standard Model also is left for the future.

\newsec{Acknowledgements}

We want to thank J.L.Cardy, P.H.Damgaard, G.Shore and A.B.Zamolodchikov
for useful discussions. This work has been supported by CYCIT
and the EEC Science Twinning grant SC1000337. X. V. acknowledges
the support of an FPI grant.

\vfill\eject


{\bf Appendix - Conformal Perturbation Theory in $d$-Dimensions}
\bigskip \bigskip
Conformal invariance fixes in {\sl any} dimension the form of the two-
and three-point correlators.
This property can be used to set a perturbative expansion around
conformal field theories which are non-Gaussian (See also ref.\mavro).
Standard perturbation
theory appears as a particular case of this conformal perturbation
theory.
\bigskip
Let us consider, then, a quantum field theory, invariant under the
conformal group, in an arbitrary dimension $d$. We denote by
$\phi_i \left( x \right)$ a generic field and by $\Delta_i$ its
dimension. Conformal invariance fixes the form of the two- and three-point
 functions to be
\eqn\ttpf{\eqalign{{\vev{ \phi_i \left( x \right) \phi_j
\left( y
\right) }}_{CFT}&=  {\delta_{i j} \over  \left\vert x -y \right\vert^{2
\Delta_i}} \cr{
\vev{ \phi_i \left( x \right) \phi_j \left( y \right) \phi_k
\left( z \right) }}_{CFT} &= { C_{i j k} \over
 \left\vert x - y
\right\vert^{\Delta_i + \Delta_j - \Delta_k}  \left\vert y - z
\right\vert^{\Delta_j + \Delta_k - \Delta_j}  \left\vert x - z
\right\vert^{\Delta_i + \Delta_k -\Delta_j}},}}
where $C_{ijk}
$ are constant coefficients called structure constants.

A conformally invariant field theory can be perturbed with one of its
operators (call it $\phi_p$, with dimension $\Delta
_p$). A correlator in the
perturbed theory is defined to be
\eqn\cpc{ \vev{ \phi_1 \left( x_1 \right) \dots \phi_N \left( x_N
\right) }
 = { \vev{ \phi_1 \left( x_1 \right) \dots \phi_N \left( x_N \right)
e^{\displaystyle \lambda_0 \int d^d x \phi_p \left( x \right) } }_{CFT}
\over \vev{ e^{\displaystyle \lambda_0 \int d^d x \phi_p \left( x
\right) } }_{CFT}}. }
The subscript $CFT$ means that the vacuum expectation value has to be
computed in the conformal theory.
                     For two point-functions, at first order in
$\lambda_0$, formula \ttpf~ allows one to write a general expression
 for any      dimension of space-time,
\eqn\ptpf{\eqalign{\vev{ \phi \left( x \right) \phi \left( 0 \right) }
&=
\vev{ \phi \left( x \right) \phi \left( 0 \right) }_{CFT} + \lambda_0
\int d^d
y \vev{ \phi \left( x \right) \phi \left( 0 \right) \phi_p \left( y
\right) }_{CFT} + O \left( \lambda_0^2 \right) \cr
&= {1 \over \left( x^2 \right)^{\Delta
}} \left( 1 + \lambda_0 C_{\phi \phi
p} {4A\over (d-\Delta)}\vert x \vert^{d-\Delta_p} +
 O \left( \lambda_0^2 \right) \right). \cr}}
The constant $A$ is given by
\eqn\b{ A =  { \pi^{d \over 2} \Gamma \left(
\Delta_p - {d \over 2} \right) \Gamma^2 \left(1+ {d-\Delta_p\over 2}
 \right)
\over \Gamma^2 \left( {\Delta_p\over 2}\right)
 \Gamma \left(1+ d - \Delta_p \right)}.}
Remark that, for $\left( d - \Delta_p \right)
 \to 0$, $A$ goes to $  \pi^{d\over 2}$.
 This is the case of slightly
relevant perturbing fields. For these nearly marginal perturbations,
expression \ptpf~ does not make sense and needs renormalization.

At this
point, the procedure mimics the two-dimensional case \zamcth \cl
, so we will skip
the details, recalling only some important points. We define the
renormalized field and the renormalized coupling constant as
\eqn\fieldrenorm{ \Phi \left( x , g \right) \equiv
 {1 \over \sqrt{Z}} \phi
\left( x \right),\qquad g\equiv Z_g \lambda_0\quad.}
We then set
            wave function and coupling constant renormalizations
                 at a scale $\kappa$ to be
\eqn\wavefctrenorm{ \left.
\vev{ \Phi_p \left( x , g \right)  \Phi_p \left( 0 ,
g \right) }
\right\vert_{\vert x \vert = \kappa^{-1}} = \kappa^{2d}.}
\eqn\teta{ \Theta \left( x \right) = V_d \beta \left( g \right) \Phi_p
\left( x , g \right), }
Following the steps described in  \cl, one can get the expression
for the renormalized coupling constant in terms of the bare one
\eqn\reng{g= \kappa^{\Delta-d} \lambda_0 \left( 1 + \lambda_0
\kappa^{\Delta-d} {A\over   n-\Delta }  C_{ppp}\right)\quad.}
It is now convenient to rescale the coupling constant $g \longrightarrow
{  g\over A}$ to simplify our formulae.
Next, we compute
 the beta-function
\eqn\res{  \beta \left( g \right)  =-\left(d- \Delta_p
 \right)
 g -             C_{ppp} g^2  \quad,      }
and the anomalous dimension for the new scaling dimensions for
the perturbing field as well as for any other field
\eqn\andim{\eqalign{
 \Delta_p \left( g \right) &= \Delta_p - 2
 C_{ppp} g
 , \cr
 \Delta\left(g\right)&=\Delta-        2  C_{\phi\phi p} g.}}
{}From formulae \res, we see that there is a new fixed
point at an infinitesimal distance from the original conformal point,
\eqn\newfixpoint{g^* = - {  d-\Delta_p
 \over   C_{ppp}}\quad.}
  At this point, the scaling   dimensions are
\eqn\critanomdim{\eqalign{  \Delta_p^* &\equiv
 \Delta_p\left(g^*\right)=
2 d -\Delta_p
, \cr
 \Delta^* &\equiv
  \Delta \left( g^* \right) = \Delta+ 2 \left( d
-\Delta_p \right) {C_{\phi
\phi p} \over C_{ppp} } . \cr}}
Note that the IR scaling   dimension of the perturbing field
is                   insensitive to the value of the structure constant.
\bigskip
Using formula \res, one can also write the Callan-Symanzik equation for
the two-point correlators,
\eqn\cs{ \left(2\vert x \vert   {\partial \over \partial \vert x \vert}
+ 2 \Delta_p\left( g \right) + \beta \left( g \right) {\partial \over
\partial g} \right) \vev{ \Phi_p
 \left( x , g \right) \Phi_p \left( 0 , g
\right) } = 0, }
and find a renormalization group improved two point function for the
perturbing field as the solution of \cs~ that fulfills condition
\wavefctrenorm. Proceeding thus, one has
\eqn\rgimprovedcorr{ \vev{ \Phi_p \left( x , g \right) \Phi_p \left( 0 ,
g \right) } = { \kappa^{2d} \over \vert \kappa x \vert^{2\Delta_p}}
 { 1 \over
\left(  1 -
  g C_{ppp}{\left( \vert \kappa x \vert^{d-
\Delta_p} -
1 \right)\over d-\Delta_p}
 \right)^4}.}
With this expression, $\Delta c $ can be computed,
 using either of the following forms of the sum rule
\eqn\sumrule{ \Delta c = { d + 1 \over d V_d} \int_{\vert x\vert>\epsilon}
d^d x \vert x \vert^d \vev{ \Theta \left( x \right) \Theta \left( 0
\right) }
= \int_0^\infty d \mu {c\left( \mu \right) \over \mu^{d-2}} , }
where
 the spin 0 part of the spectral representation is
\eqn\spectralrep{  c \left( \mu \right) = {
2^d \Gamma
\left( d \right) \left( d + 1 \right) \over \pi V_d} {1 \over \mu^3}
\left.\Im {\rm m} \of{\vev{
 \Theta \left( p \right) \Theta \left( -p \right)}}
\right\vert_{p^2 = - \mu^2} \quad.    }
In any case, what we find is
\eqn\deltac{\Delta c = {8 \over 3} {d + 1 \over d}
V_d^2 { (d - \Delta_p)^3 \over
A^2 C_{ppp}^2}. }
which is the change of the $c$-charge between the UV and IR fixed points.

In general, the perturbing field does not close an algebra by itself in
the sense of the operator product expansion. This causes mixing with
other fields in the theory. It is then necessary to consider a system
of beta-functions
\eqn\betasys{\beta_i=-(d-\Delta_i) g_i -\sum_{j,k} C_{ijk}\, g_j g_k\quad,}
where $\{g_i\}$ span the coupling space. The $c$-function is then
\eqn\csys{c= -{1\over 2} \sum_i(d-\Delta_i) g_i^2 -
{1\over 3} \sum_{i,j,k} C_{ijk} \,g_i g_j g_k\quad,}
which contains the most general case. In practice, the system
\betasys~
may be simpler and solvable so that
      explicit expressions for the
variation of $c$ can be obtained, as in sect.  3.5.
\bigskip \bigskip
{\bf Example 1.}   TWO-DIMENSIONAL CONFORMAL FIELD THEORIES
\bigskip
One particular set of two-dimensional
conformal field theories is the unitary minimal
series, whose central charges are given by the formula
$          c = 1 - {6 \over m \left( m + 1 \right)}, \qquad m \ge 3,$
where $m$ is an integer.   As discussed in ref. \zamcth,
     perturbing one such minimal model with its least relevant field, if
$m$ is large, we obtain the next minimal model at $c\left( m - 1
\right)$. We can have an inkling of this fact recalling that
\eqn\largem{\eqalign{ d-\Delta_p &= {4 \over m+1}= {4\over m} +
O\left({1\over m^2}\right)\cr
C_{ppp} &= {4\over \sqrt{3}}+ O\left({1\over m}\right). \cr}}
Applying \deltac, we correctly obtain
\eqn\dcminm{ \Delta c = { 12\over m^3}+O\left({1\over m^4}
\right)= c(m)-c(m-1) ,}
which is a check for our formulae.
\bigskip \bigskip
{\bf Example 2.} THE $\eps$-EXPANSION
\bigskip
The $\eps$-expansion can be thought as an example of conformal
perturbation theory. Recall that it consists in perturbing the
massless free theory with
$\lambda \ph^4$ theory in $4-\eps$ dimensions.
In order to apply the previous formulae, we recall that
       the propagator of the scalar field                   is
\eqn\freeprop{ \vev{ \ph \left( x \right) \ph \left( 0 \right) } = { 1
\over V_d \left( d - 2 \right)} { 1 \over \vert x \vert^{d-2}}.}
To obtain   proper conformal fields, operators have to be normalized.
For instance,
$$ \phi \left( x \right) = \sqrt{ V_d \left( d - 2 \right)} \ph \left( x
\right).$$

As a conformal operator, $\ph^4$ has dimension
 $\Delta_p = 4 - 2\eps$, so
that $d-\Delta_p = \eps$.
 We see then that the operator is slightly
relevant if $\eps$ is small enough. Other conformal features are
\eqn\cfphifour{ \eqalign{ C_{ppp} &= 6^{3 \over 2} \cr
 C_{\ph \ph p}
&= 0 \cr}}
(from this last equation we see that the two point function $\vev{\ph
\ph}$ gets no correction at the first order).
This is just a particular case of the general conformal perturbation
expansion and it is elaborated in sect.  3.4.
\vfill\eject
\listrefs

\newsec{Figure Captions}
\bigskip
\parindent=0pt
{\bf fig. 1}

RG pattern which  exemplifies the additivity property of the
$c$-charge.

\bigskip
{\bf fig. 2}

Schematic picture of two RG flows leading to the purely massive
phase which are not deformable  into each other.

\bigskip
{\bf fig. 3}

RG flows for the $r$-th and ($r-1$)-th Ginsburg-Landau models between
two and four dimensions.

\bigskip
{\bf fig. 4}

Chains of RG flows for the multicomponent $\varphi^4$ theory
(see text): \quad a) Gaussian $\longrightarrow$ Symmetric $\longrightarrow$
$(x^*,y^*)$, for $N>4$; \quad b) Gaussian $\longrightarrow$ Symmetric
$\longrightarrow$ Decoupled, for $N>10$.

\bigskip
{\bf fig. 5}

Level map of the $c$-function for the  multicomponent $\lambda \varphi^4$
theory with $N=8$ and $d=3.8$.

\bigskip
{\bf fig. 6}

Schematic picture of the RG flows in the linear and non-linear
sigma-models.

\vfill\end